\documentclass[a4paper,amsmath,amssymb,twocolumn]{revtex4} % Per la
\usepackage{graphicx} \input{epsf}

\newcommand{\vect}[1]{\mathbf{#1}}
\newcommand{\tdot}[1]{{\hskip2pt\ddot{\null}\hskip2.5pt
\dot{\null}\kern -5pt {#1}}} \newcommand{\alfa}{\alpha}
\newcommand{\passo}{l}
\newcommand{\raggio}{a}
\newcommand{\omegac}{\omega_c}
\newcommand{\omegap}{\omega_p}
\newcommand{\massae}{m_e}
\begin{document}

\title{A first principles explanation for the density limit in
magnetized plasmas}

\author{M. Zuin$^{1}$, A. Carati$^{2}$, M. Marino$^{2}$, E. Martines$^{1}$ and L. Galgani$^{2}$}
\affiliation{$^{1}$Consorzio RFX, Associazione EURATOM-ENEA sulla Fusione,
Padova, Italy\\$^{2}$Dipartimento di   Matematica, Universit\`a degli
Studi di Milano, Milano, Italy}

%\author{M. Zuin$^{1}$, E. Martines$^{1}$}
%\affiliation{$^{1}$Consorzio RFX, Associazione EURATOM-ENEA sulla Fusione,
% 35127 Padova, Italy\\$^{2}$Dipartimento di Fisica 'G. Galilei', %Universit\`{a} degli Studi di Padova,  35131 Padova, Italy}

\date{\today}

%\end{abstract}
\maketitle \textbf{Fusion research on magnetic confinement is
confronted with a severe problem concerning the electron densities $n_e$
to be used in fusion devices. Indeed, high
densities are mandatory for obtaining large efficiencies,
whereas it is empirically found that catastrophic disruptive events
occur for densities exceeding a maximal one $n_e^M$.  On the other hand,
despite the large theoretical work  
``there is no widely accepted, first principles model  
for the density limit'' (see \cite{green2}, abstract).  
Here, we propose a simple microscopic
model of a magnetized plasma suited for a tokamak, for which   the
existence of a density limit is proven. This property 
turns out to be a general  collective feature of 
electrodynamics of point charges, which is lost in the continuum 
approximation. The law we find is
\begin{equation} \label{legge}
n_e^M = 1.74\, \frac{1} {\massae c^2}\, \frac{B^2} {\mu_0}
\end{equation}
where  $\mu_0$ is the vacuum  permeability, $c$  the speed of light, $\massae$
 the electron mass,  and $B$ the  magnetic  field.
As shown in Fig. 1,  the theoretical limit (big circles) is 
in rather good agreement with the empirical data, actually a
 surprisingly good one for a model based on first principles, 
with no adjustable  parameter. }

The way in which law (\ref{legge}) was established is an
interesting example of an encounter between fundamental and
applied research. Indeed three of the present authors are involved 
since some time in studies of a general character concerning
the  microscopic electrodynamics of systems of point
particles (see \cite{cg} and \cite{mcg}), in which both
the mutual retarded forces, and the
radiation reaction force of Abraham Lorentz and Dirac  \cite{dirac}
(see also \cite{massimodirac} and \cite{jackson}) are taken into account. 
One of the  results obtained is the proof of an
identity conceived by Wheeler and Feynman \cite{wf}, 
and an appreciation of the
role the latter plays in allowing for the very existence of a
dispersion relation. In particular, some examples of dispersion
relations were given (see \cite{cg}, Fig. 1),  which exhibit,  as the matter density is increased, a bifurcation of a topological
character, entailing an instability.
But the physical relevance of this fact was not emphasized. 
Such a density controlled bifurcation impressed instead very much those of the present authors who deal with plasma physics, who suggested it may be  relevant  for fusion plasmas. 
To this end, the simplest possible model was formulated, 
that should capture, within the frame of the foundational
works mentioned, the
essential  physics of a magnetized plasma, confined in a tokamak configuration (the most studied one for fusion plasmas). 
The model is presented here, together with the deduction of law
(\ref{legge}). Preliminarily, the main evidence for the existence of a  density
limit in tokamaks is recalled, and it is discussed how well does law
(\ref{legge}) fit the data.

For the purposes of the present paper, all is needed 
to know about tokamaks is essentially that they are toroidal devices in which
the confining magnetic field $B$ is the vector sum of  a 
strong toroidal field  $B_t$ produced by a
set of coils wound around a torus, and of a much smaller poloidal field $B_p$ 
generated by a toroidal plasma current $I_p$. A few more details will be mentioned later.

\begin{figure}
\includegraphics[width=1.\columnwidth]{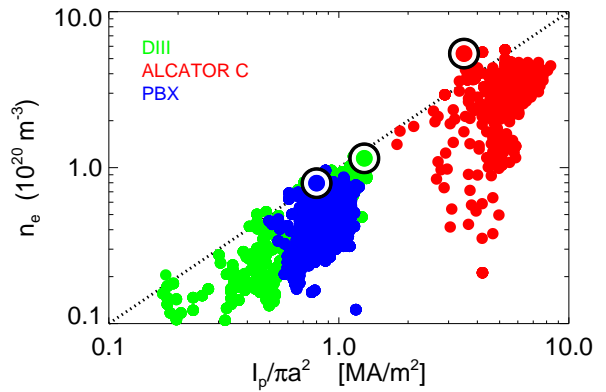}
\caption{Greenwald plot for three tokamak devices (data extracted from \cite{green2}). Dotted line is the empirical Greenwald limit (\ref{legge_greenwald}). Big circles are the theoretical predictions (\ref{legge}) for the same devices. }
\label{fig:1}
\end{figure}

The empirical data which show the existence of a density threshold
beyond which tokamaks cannot operate, were collected by
Greenwald \cite{green2} 
in a classical figure, the data of which
are reported here in Fig.1.  In the figure, the 
electron densities  $n_e$ at which
three different tokamak devices could
actually be operated are reported  versus the so-called Greenwald parameter
$n_G \equiv I_p/\pi \raggio^2$ (where $\raggio$ is the minor radius of the torus), which is presumed to be the relevant control parameter.
Indeed,  the law proposed by Greenwald for the maximal density (dotted line in the figure) is  
\begin{equation} \label{legge_greenwald}
n_e^M = \alpha_G\, I_p/\pi\raggio^2 \ ,  
\end{equation}
where $\alpha_G$ is a 
constant with suitable dimensions, such that  
 $\alpha_G = 1$ in the units  indicated in the figure. 
 The theoretical predictions given by (\ref{legge}) are also reported as big circles.
\begin{figure*}[t]
\centering \includegraphics[width=.7\columnwidth,angle=90]{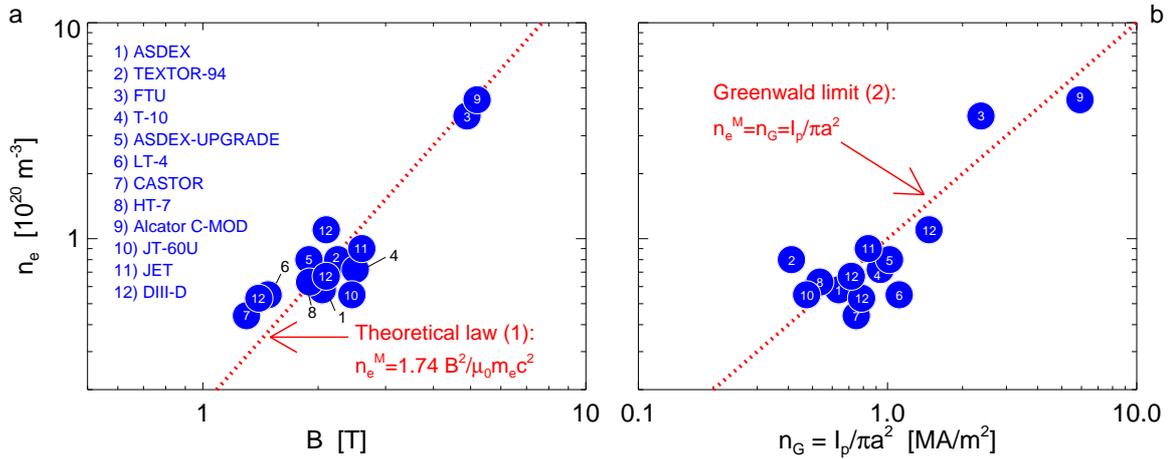}
\caption{Density limit values for various conventional tokamak machines \emph{vs}  $B$ in \textbf{a} and \emph{vs} $I_p/\pi a^2$ in \textbf{b}. Dotted line in \textbf{a} is the proposed theoretical law (\ref{legge}), and in \textbf{b} is the empirical law (\ref{legge_greenwald}).  For the various devices the points given are representative of disruptive events actually due to an increase of density  or of an operative condition declared to be close to a density limit (see references \cite{Stabler,DeVries,Frigione,Merezhkin,Merthens,Howard,Dyabilin,Asif,Bombard,Takenaga,Saibene,Petrie}).} \label{fig:2}
\end{figure*}

Thus there naturally arises the  question, how is it possible that two analytically different  predictions, (\ref{legge}) and (\ref{legge_greenwald}), happen to agree with each other, at least in a few definite cases. The reason is that one has 
\begin{equation}\label{relazione}
\frac{B}{\mu_0} \approx \, b \, \frac{I_p}{\pi a^2} \ , \quad b\,=\,q(a)\, R\, \frac{1}{\kappa^{1/2}(1+\kappa^2)} \ ,
\end{equation}
where the dimensional coefficient $b$ (a length) is not a universal
one, but  depends parametrically on geometric factors and operative 
conditions  characterising each experiment. These are
the major and the minor  radii $R$ and $a$, the plasma elongation $\kappa$ (equal to 1 for
circular plasmas, see \cite{Miyamoto} pg. 277), and the edge safety 
factor $q(a)$ defined below.

This is seen as follows. 
One has $B_p=\mu_0 I_p/(2 \pi a \kappa^{1/2})$.   
Furthermore, in the approximation $B \approx B_t$, one has 
$B \approx (B_t/B_p)B_p$, while the operative parameter $B_t/B_p$ 
is determined by the edge safety factor $q(a)$ which, 
for the simplified case of an elliptical plasma (see \cite{Miyamoto}), 
is defined as
$q(a)=  (a/R) (B_t/ B_p)  (1+ \kappa^2)\, /2$. This gives relation
(\ref{relazione}). The formula for  $b$ in the general case is also 
easily established.

Now, magnetohydrodynamic stability requires $q(a)>2$, but the actual 
value at which each experimental data point of Fig. \ref{fig:1} was
taken is not given in the literature. So we  assumed $q(a)=4$, 
which is a typical operational value, and  this introduces an uncertainty in the theoretical points reported in Fig. \ref{fig:1}. % reported in the figure. 

Thus, we decided to look directly at the experimental values available
in the literature, from which a definite estimate of $B$ could be
obtained, and this we did
for conventional tokamaks more recent than those in Fig. \ref{fig:1}.   
Such values are plotted in Fig. \ref{fig:2}, where they are compared to  laws (\ref{legge}) and (\ref{legge_greenwald}). 
The agreement with law (\ref{legge}) is  perhaps a little better. %  both for low and high magnetic field tokamaks. %Notice that,  at variance with the data of Fig. 1, which refer to accessible density regimes, and so in general just give an estimate of the threshold from below, the data of  Fig.2 refer instead to  cases in which a disruption actually occurred due to an increase of the density.

This fact  might have relevant  implications for future tokamaks, as
it implies a favorable density scaling for machines with large values
of the product 
$R\,  B$. For example, let us consider the international thermonuclear  
experimental reactor (ITER) \cite{Iter},  which should operate at a 
toroidal field of 5.3 T. According to  law (\ref{legge}) it would be 
able to operate  at densities up to $n_e^M= 4.7\times 10^{20}
m^{-3}$. This is a value  more than three times larger than that 
expected according to  Greenwald law (\ref{legge_greenwald}) for a plasma current of 15
MA, which is the corresponding value  of $I_p$ for a $q(a)=3$ scenario.

It is worth mentioning that a $B^2$ dependence of the density limit in the ALCATOR C experiment was noted in the past by Granetz \cite{Granetz}, although such a clear dependence was not observed on other experiments (see for example \cite{Petrie} for the DIII-D tokamak).
Also, one should point out  that tokamaks with very low aspect 
ratio $R/a$ (spherical tokamaks) and reversed field pinches (RFP) \cite{Lorenzini}, for which the validity of the Greenwald scaling has been proposed \cite{Valisa},  seem not to fit well into the
proposed $B^2$ scaling. This,  perhaps, suggests 
that at low magnetic fields other effects,  not considered in the present simple model, might come into play.
  
We show now  how law (\ref{legge}) was obtained, in
the frame of microscopic electrodynamics of point particles (see \cite{cg} and \cite{mcg}),
rather than of  
magnetohydrodynamics (see for example
\cite{gold}, chapter 17), or of the mean field theories of the
Vlasov approach. We describe 
the plasma as constituted of  point particles obeying Newton
equations, with  both the
retarded electromagnetic interactions among all particles and the radiation reaction force
taken into account. 
We then concentrate on the role played by the gyration
of the electrons around the magnetic field lines, and so ignore their
motions along the field lines, and also ignore the electrostatic part
of the problem. Finally we also limit ourselves
to the extremely simplified case of a one--dimensional array.

So we introduce the following model.  Given a constant magnetic field
$B$, which we take oriented along the ${z}$ axis of a
cartesian coordinate system, we constrain each electron, say the $n$-th
one, to move on a plane parallel to the ($x,y$) plane, 
so that its $z_n$ coordinate is fixed. The simplest choice is to
take $z_n=n \passo$, with $n \in Z$, for a given positive step
$\passo$. 
Each electron, say the $n$--th one,  
is subjected to the external
magnetic field $B$, and also    to the electromagnetic field created by  the
electrons themselves. %The latter field can   be split into two contributions.  
Namely, the sum of the
 Li\'enard--Wiechert fields, which are
determined as the retarded  
solutions of Maxwell equations having as sources the charge and the current
densities  of each other  electron $m\neq n$, and 
 the Abraham--Lorentz--Dirac  radiation reaction force, due
to the  motion of the $n$--th electron itself, The latter is given,  in the
nonrelativistic approximation, by $(2/3) e^2/(4\pi \epsilon_0
c^3)\,  \tdot{\vect x}_n$,
 where $e$ is the electron charge 
and ${\vect x}_n$ the position vector of the electron. 
 
We then perform the dipole approximation. Thus  
we neglect the magnetic field due to the $m$--th electron, and  for
the electric field  created by it we take the well known 
expression for a  dipole. 
Finally, we approximate the distance between
electrons  $n$ and $m$
by $r_{n,m}=\passo |n-m|$. 
The system of equations of motion defining the
model  is then
\begin{widetext}
\begin{equation}
\begin{split}
{\ddot x}_n- \omegac {\dot y}_n -\frac 23\, \frac{e^2 }
{4\pi\epsilon_0 \massae c^3}\, {\tdot x}_n &= -
\frac{e^2 }{4\pi\epsilon_0 \massae } \sum_{m\neq n}\Big[ \frac{x_m (t-r_{nm}/c)}{r^3_{nm}} +
\frac{1}{c}\frac{{\dot x}_m (t-r_{nm}/c)}{r^2_{nm}} + \frac{1}{c^2}
\frac{{\ddot x}_m(t-r_{nm}/c)}{r_{nm}}\Big]\\ {\ddot y}_n+ \omegac
{\dot x}_n -\frac 23\, 
\frac{e^2 }{4\pi\epsilon_0 \massae c^3}\, {\tdot y}_n &= - 
\frac{e^2 }{4\pi\epsilon_0 \massae } \sum_{m\neq
n}\Big[ \frac{y_m(t-r_{nm}/c)}{r^3_{nm}} + \frac{1}{c}\frac{{\dot
y}_m(t-r_{nm}/c)}{r^2_{nm}} + \frac{1}{c^2} \frac{{\ddot
y}_m(t-r_{nm}/c)}{r_{nm}}\Big]\\
\end{split}
\end{equation}
\end{widetext}
for $n\in Z$, where $\epsilon_0$ is the vacuum permittivity, and
 $\omegac=eB/\massae$  the Larmor or cyclotron frequency of the
 electrons in the external magnetic field $B$. This is an infinite
 system of linear equations with delay, which is just a simple variant
 of the system considered in \cite{cg}.

Our aim is now to investigate the stability properties of the system,
as the control parameters  $n_e=1/\passo^3$ and $B$ (or
equivalently $\omegac$) are varied. 
Following a completely standard procedure 
(see for example  \cite{chandra}), we 
compute the normal
modes of the system and determine the values of the parameters for
which the frequencies become complex.  So we look for normal mode
solutions with wavenumber $k$ and angular frequency $\omega$, i.e.,
of the form
\begin{equation}
x_j=A_x e^{i(k\passo j+\omega t)}\ , \quad y_j=A_ye^{i(k\passo j+\omega t)}\ .
\label{}
\end{equation}
This leads to a linear system in
the unknowns $A_x$, $A_y$, from which the dispersion relation between
$\omega$ and $k$ is found by equating the determinant to zero. 
This gives  two real equations in the two
unknowns $\omega$ and $k$, namely,
\begin{equation}\label{eq:4}
 \Big(\frac {\omega}{\omegac}\Big)^2  \pm \, \frac\omega\omegac \ + p\, 
F(k\passo,\passo\omega/c) = 0 , 
\quad p= \frac{\omegap^2}{\omegac^2}\ . 
\end{equation}
\begin{equation}\label{eq:2}
\frac 23\, \frac{e^2}{4\pi\epsilon_0 \massae c^3} \, \omega^3 - 
G(k\passo ,\passo\omega/c) = 0\ .
\end{equation}
Here, $\omegap$ is the familiar plasma frequency defined by 
\begin{equation}
\omegap^2=e^2/\epsilon_0 \massae\passo^3\ =n_e \, e^2/\epsilon_0 \massae \ ,
\end{equation}
 while $F$ and $G$,
as functions of the variables $\alpha=k\passo$, $\beta=\passo\omega/c $, are
defined by
\begin{equation*}\label{eq:3}
\begin{split}
F(\alpha,\beta) &=\frac 1{4\pi} \Big[ \beta^2\log \big(2\,
  |\cos\beta-\cos\alpha| \,
\big)- f(\alpha,\beta)\Big]\\ G(\alpha,\beta)&=\beta^3 -g(\alpha,\beta) \ ,
\end{split}
\end{equation*}
the functions $f$ and $g$ being the ones already  introduced in 
\cite{cg}, namely,
\begin{equation*}
\begin{split}
f(\alfa,\beta)=\sum_{n\neq 0}(\frac{\cos(n\alfa-|n|
  \beta)}{|n^3|}-\beta \frac{\sin(n\alfa-|n| \beta)}{|n^2|}) \\
g(\alfa,\beta)=\sum_{n\neq 0}(\frac{\sin(n\alfa-|n| \beta)}{|n^3|} +
  \beta \frac{\cos(n\alfa-|n| \beta)}{|n^2|})\ .
\end{split}
\end{equation*}
Some details concerning the summation of the series leading to  the
term $\beta^2\log \big(2\, |\cos\beta-\cos\alpha|\big) $ entering the 
function $F$ are here omitted.

Now, one meets here with a deep question of principle. Indeed, for
fixed values of the parameters $\passo$ and $\omegac$ one has two
equations in two unknowns ($\omega$ and $k$), and this would not allow for
the existence of a dispersion relation, i.e.,  of a 
function $\omega=\omega(k)$ for a continuous  range  of values of $k$.
However, the existence  of a dispersion relation  is guaranteed by the 
fact that equation  (\ref{eq:2}) actually is  an identity.
In fact, this is a particular case of a general
identity, conceived by Wheeler and Feynman \cite{wf}  and first proven in  \cite{cg} (see section 6) for a one-dimensional case and in \cite{mcg} for a three-dimensional one.

\begin{figure}
\centering \includegraphics[width=1.\columnwidth]{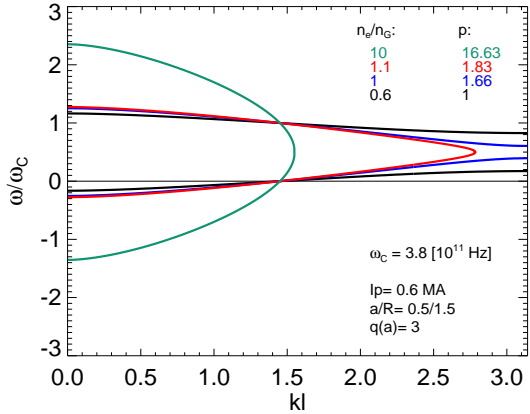}
\caption{The dispersion curves, solutions of equation (\ref{eq:4}) (with the plus sign), in the plane ($k\passo$, $\omega/\omegac$)
  for  $\omegac=3.8 \times 10^{11}$ Hz, and
for several values of the parameter $p=\omega_p^2/\omegac^2$ 
(or equivalently of the electron density $n_e$, normalized to
the empirical $n_G$ limit).} \label{fig:3}
\end{figure}

So, the problem of obtaining the  dispersion relation
is reduced to solving  (\ref{eq:4}) in the unknown 
$\omega=\omega(k)$, in which $p=\omega_p^2/\omega_c^2$ plays 
the role of a parameter.
In Fig. \ref{fig:3} the dispersion relations are shown 
for a  cyclotron frequency $\omegac=3.8\cdot 10^{11} Hz$,
and for several values of the parameter $p=\omega_p^2/\omegac^2$ (or
of the  corresponding electron density  $n_e/ n_G$, normalized to the
Greenwald density $n_G\equiv I_p/\pi \raggio^2$). 

The most important qualitative result is that normal
modes  are found to exist (for all $k$) only below a critical value of $p$, i.e., below a certain threshold 
of plasma density. Indeed,
starting up from low densities, at a certain critical density a
bifurcation is seen to occur, characterized by the fact that the curves no more
intersect the  vertical axis $k\passo=\pi$. This means that  
for values of $k$ just below $\pi/\passo$
 equation (\ref{eq:4})
does not admit a real solution, so that the
corresponding frequencies acquire an imaginary part, and 
the whole system becomes unstable. Numerical computations not reported here show that the characteristic time of the instability is of the order of $2\pi/\omegac$ and that above the critical density the Wheeler and Feynman identity is no more satisfied.

Notice that this phenomenon of
the existence of a maximal allowed density is
obviously lost if one introduces the continuum approximation, i.e., is a characteristic feature of the discrete structure of matter. Indeed,
following \cite{mcg}, the continuum approximation corresponds to deal with
wavelengths much larger than the step $\passo$, i.e., 
to  assume   
$k\ll \pi/\passo$, whereas the existence of a density limit depends on
the behavior of the system for  $k\passo \simeq \pi$.
%So the existence of a density limit turns out to be a characteristic feature of the discrete structure of matter. 

We have now to determine the
bifurcation value of the parameter $p$.  As the bifurcation occurs
for $k\passo=\pi$ and for values of $\omega/\omegac\leq 1$, i.e.,
for $\passo\omega/c\simeq 0$, one can just limit oneself to study
equation (\ref{eq:4}) for  a fixed value of the function $F$,
namely $F(\pi, 0)$, so that one is simply reduced to deal
with an algebraic equation of second degree. One computes $F(\pi,
0)\simeq0.14$, and so  real values of $\omega$ are found to exist only 
for $p\lesssim1.74$. This, together with the definition
of $p$ in (\ref{eq:4}) and $\epsilon_0 \mu_0=1/c^2 $, 
gives  law (\ref{legge}).

Notice that law (\ref{legge}) has the same form of the Brillouin limit 
\cite{Brillouin}, which is known to apply to the case of nonneutral 
plasmas \cite{Davidson}. The main difference with respect to our
procedure is that in the case of the Brillouin limit  the electric
field acting on each electron is introduced within a mean field
approach, whereas here it is computed
 in the frame of a many--body microscopic theory. Correspondingly, we
find that the instability involves normal modes with
wavelengths of the order of the mean electron distance, so that it
escapes a mean field approach. 
In particular, such an instability is found to occur  in neutral plasmas,
for which the mean charge density vanishes, 
and the Brillouin approach cannot be used.

A final comment concerns the possibility of dealing with the other main
magnetic configuration studied for the confinement of fusion-relevant
plasmas, i.e., the Stellarator \cite{Boozer}.  The present model does not directly apply. Indeed,  in the Stellarator a large amount of
power is typically transferred to the electrons through electron
cyclotron resonance heating (ECRH), and this requires to add in our
model a forcing term. 

In conclusion, through an extremely simplified model of a magnetized
plasma suited for a tokamak, based on first principles, 
we have proved the existence of a 
density limit, beyond which the system becomes unstable. The law thus 
found differs from the usually accepted one, and this  fact might 
have relevant implications for future tokamaks.

%, as it implies a favorable density scaling for machines with large values of the product ($R \cdot B$). For example, the international thermonuclear  experimental reactor (ITER) \cite{Iter},  which should operate at a toroidal field of 5.3 T,  according to  law (\ref{legge}) would be able to operate  at densities up to $n_e^M= 4.7\times 10^{20} m^{-3}$, which is a value  more than three times larger than the one expected according to the  Greenwald law for a plasma current of 15 MA, which is the corresponding value  of $I_p$ for a $q(a)=3$ scenario.

%Finally we may mention that preliminary results seem to indicate that a law of the form (\ref{legge}) should hold also in a more realistic three dimensional model, in which the role of the ions too is taken into account.

\emph{The authors wish to thank Dr. Nicola Vianello for fruitful  discussions. \\
This work, supported by the European Communities
  under the contract of Association between EURATOM/ENEA, was carried
  out within the framework the European Fusion Development Agreement.}

%\clearpage

\end{document}